# Analyzing dynamical disorder for charge transport in organic semiconductors via machine learning

*Patrick Reiser[1,2], Manuel Konrad[1], Artem Fediai[1], Salvador Léon[3], Wolfgang Wenzel[1] and Pascal Friederich[1,2,]\**

[1]*Institute of Nanotechnology, Karlsruhe Institute of Technology (KIT), Hermann-von-Helmholtz-Platz 1, 76344 Eggenstein-Leopoldshafen, Germany*
[2]*Institute of Theoretical Informatics, Karlsruhe Institute of Technology (KIT), Am Fasanengarten 5, 76131 Karlsruhe, Germany*
[3]*Department of Industrial Chemical Engineering and Environment, Universidad Politécnica de Madrid, C/ José Gutierrez Abascal, 2, 28006 Madrid, Spain*

**Abstract:**
Organic semiconductors are indispensable for today's display technologies in form of organic light emitting diodes (OLEDs) and further optoelectronic applications. However, organic materials do not reach the same charge carrier mobility as inorganic semiconductors, limiting the efficiency of devices. To find or even design new organic semiconductors with higher charge carrier mobility, computational approaches, in particular multiscale models, are becoming increasingly important. However, such models are computationally very costly, especially when large systems and long time scales are required, which is the case to compute static and dynamic energy disorder, *i.e.* dominant factor to determine charge transport. Here we overcome this drawback by integrating machine learning models into multiscale simulations. This allows us to obtain unprecedented insight into relevant microscopic materials properties, in particular static and dynamic disorder contributions for a series of application-relevant molecules. We find that static disorder and thus the distribution of shallow traps is highly asymmetrical for many materials, impacting widely considered Gaussian disorder models. We furthermore analyse characteristic energy level fluctuation times and compare them to typical hopping rates to evaluate the importance of dynamic disorder for charge transport. We hope that our findings will significantly improve the accuracy of computational methods used to predict application relevant materials properties of organic semiconductors, and thus make these methods applicable for virtual materials design.

## 1. Introduction

**Motivation.** Organic semiconductors have been successfully integrated into many commercially relevant electronic devices such as in organic light-emitting diodes (OLEDs),[1] organic photovoltaics (OPV),[2] organic lasers,[3] and organic transistors (OFETs)[4]. Their optoelectronic properties and potential flexibility, role-to-role manufacturing[5] or biocompatibility[6] opens many interesting applications. However, concerning charge carrier transport and mobility, organic semiconductors still fall short behind their inorganic counterparts. Electrochemical doping of organic semiconductors does significantly increase conductivity as well as charge carrier injection but can reduce long-term stability and efficiency due to drift and diffusion of dopants.[7–10] Although organic crystals achieve high mobilities in the range of 1-100 $cm^2$/Vs,[11,12] they are usually not used in OLEDs. An OLED made from a single crystalline layer will likely suffer from charge carrier injection and recombination imbalances.[13] To achieve isotropic light emission and homogeneous layer thickness in device stacks with multiple layers, amorphous or semicrystalline materials are required, having significantly lower charge carrier mobilities. Consequently, it is crucial to understand and to improve charge transport, in particular charge mobility, for common OLED and OPV materials, which is determined primarily by the width of the disorder distribution, as shown by Bässler et al.[14–16] The enormous suppression of the mobility with increasing disorder is largely responsible for the poor performance of amorphous organic semiconductors with respect to mobility. Multiscale modeling methods had large success in *ab-initio* mobility prediction for various materials and therefore serve as a screening tool for new or modified molecules.[7,17–24] In most models, charge transport is described as a sequence of hopping processes, *i.e.*, thermally assisted tunneling of localized polarons between neighboring molecules.[25] Hopping rates can be estimated based on the Miller-Abrahams model and Marcus theory,[26,27] which require reorganization energies and transition matrix elements from ab-initio DFT calculations, as well as the spatial distribution of molecular energy levels. The hopping rates can then be inserted into master equations,[28–30] Monte Carlo simulations[31] or mean field theories[32] to determine the carrier mobility of interest.[33,34]

Presently most quantitative models for the mobility assume a static picture with frozen molecules and thus time-independent energy levels, although molecular morphologies are often created by dynamics simulations.[21,35,36] This is partly due to the numerical effort of electronic structure methods, such as DFT, which precludes their use in large systems with full dynamics. For this reason, the role of dynamic disorder is presently not well understood.

**Static and dynamic disorder.** In amorphous molecular solids, at any given point in time, the distribution of orbital energies across all molecules in the system is typically estimated to be Gaussian with a material-dependent standard deviation, which is referred to as (global) energy disorder in the Gaussian disorder model.[14] The disorder parameter σ constitutes a sensitive factor for mobility prediction: $\mu \propto e^{-C(\sigma/kT)^2}$. However, due to thermal vibrations around the individual (and potentially different) equilibrium conformations of each molecule in the amorphous matrix, these energy levels are fluctuating in time. The thermally driven conformational disorder can include methyl rotations,[37] pedal motions,[38] resonances in hydrogen bonds[39] or dipole orientational[40] disorder.[41,42] This fluctuation is referred to as dynamic disorder,[43–45] which is related to electron-phonon interactions which cause a spread of energy levels. If all molecules had the same equilibrium conformation, all time-averaged energy levels would be equal. However, in most amorphous materials, the molecules are constrained in individual equilibrium structures, leading to a spread of the time-averaged energy levels, which is referred to as static disorder.[46] It will depend on the size and flexibility of the molecules, on the response of energy levels to conformational changes, on the degree of crystallinity and on possible additives such as dopants.[47] The effective energy landscape and energy disorder experienced by the charge carriers thus depends on the typical residence time of charge carriers,[48,49] *i.e.*, electrons and holes localized on a single molecule. In case the residence time is much smaller than the fluctuation time of energy levels (high electronic coupling between molecules), the dynamic disorder has to be taken into account, and the total disorder becomes an upper bound of the energy disorder. However, in case of lower hopping rates, each charge carrier will experience a fluctuating energy landscape potentially with multiple resonances with neighbouring energy levels, leaving the static disorder as a lower bound.[48] As a consequence, the thermal energy fluctuations and the resulting dynamic disorder can be beneficial or detrimental to charge transport. This will depend on whether the charge transport is limited by either static disorder or electronic couplings between neighbouring molecules, where the transfer integral is influenced by orientation and distance as well. Additionally, electrostatic interaction will modify the energy landscape, as soon as a charge is moving in the system. Depending on the transport model, the energy distribution integrated over in percolation path theory for mobility prediction or used in Monte-Carlo simulations to initialize site energies, is often assumed to be static. However, the disorder is commonly estimated by the total disorder, or the morphologies are generated by dynamics simulations only, because of the computational cost of energy calculations.[21,33,35,36,50]

Therefore, we incorporated faster machine learning (ML)[51] models to upscale energy predictions on long trajectories with thousands of molecules in order to obtain insights into the dynamics of disorder effects and to distinguish static and dynamic disorder contributions.[43,44,52–54]

**This work.** In a recent proof of principle study, we showed that energy level fluctuations in PEDOT:PSS can be evaluated using machine learning models, enabling the calculation of dynamic and static disorder in PEDOT oligomers.[55] In this paper, we present an extended study of static and dynamic disorder, evaluated using ML models for OLED and OPV application relevant materials (depicted in **Scheme 1**). For this study, we focus on conformational disorder and evaluate vacuum single point energies for isolated molecules from DFT reference calculations to train the ML models. Electrostatic disorder due to the electrostatic potential of neighbouring molecules as well as additional electrons and holes is neglected here, but plays an important role to quantify charge carrier mobility in amorphous organic semiconductors.[21,34] We show that machine learning models can be trained to reliably predict orbital energies with a limited number of DFT calculations and subsequently used to compute static and dynamic disorder on the full set of MD trajectories. Thereby we obtain good statistics for a comprehensive analysis of energy disorder as a function of time and position. This enables us to compare dynamic energy changes between neighboring molecules with typical hopping rates of charge carriers. In this work, we have succeeded in separating the static disorder from the dynamic disorder, which is often larger than the former. The comparison of hopping and energy level fluctuation rates is required to verify the static disorder being used for multiscale models. Nonetheless, we discuss that for a certain parameter range the dynamic disorder could become relevant to more accurately predict the charge carrier mobility of amorphous organic semiconductors. We expect that our method will enable accurate prediction of the charge carrier mobility in amorphous organic semiconductors, paving the way to a reliable virtual design of high-mobility amorphous organic semiconductors.

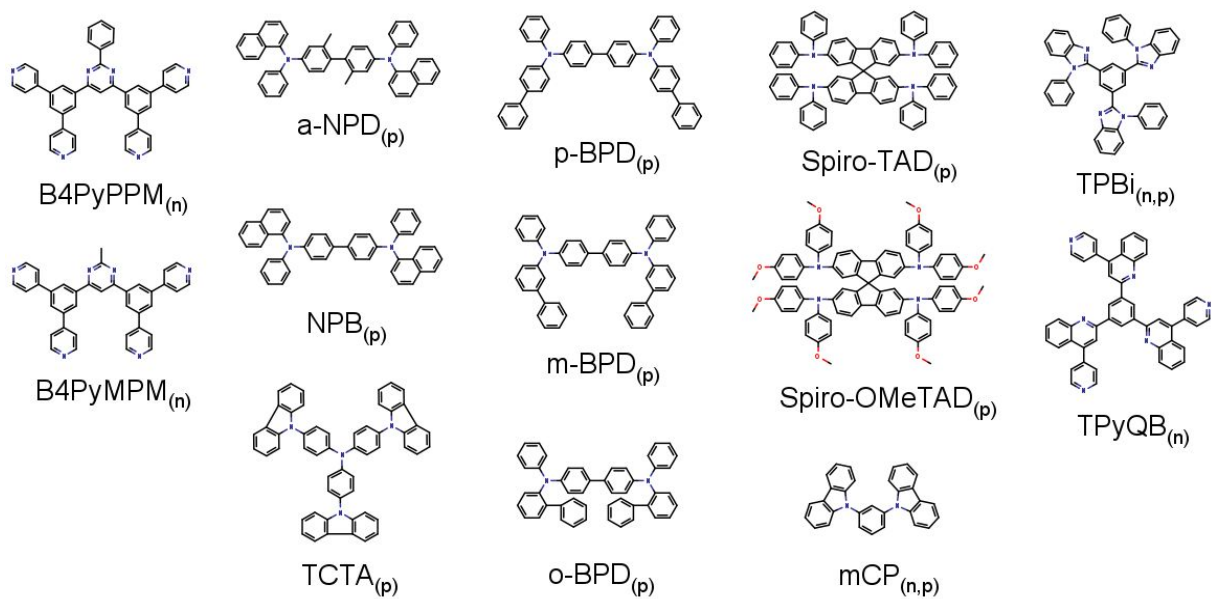

**Scheme 1: Molecules.** *Structure for all molecules investigated in this study. They can be categorised into p-type and n-type semiconductors. P-type: a-NPD, NPB, Spiro-TAD, Spiro-OMeTAD, TCTA, BPD and mCP. N-type: B4PyPPM, B4PyMPM, TPBi and TPyQB.*

## 2. Methods

The workflow used in this study consists of the parameterization of molecule specific classical force fields (FF), molecular dynamics simulations (MD) of the amorphous solid, DFT calculations on individual snapshots to generate a sufficiently large dataset of HOMO energy levels, training of machine learning models, in particular neural networks (NN), to predict energy levels as a function of the conformation of a molecule, and finally energetics prediction on the full MD trajectories using the ML model instead of expensive DFT calculations. The individual steps of the workflow are depicted in a schematic overview of **Figure 1**. In the following, we will give information on each step and their mutual relation. With this procedure all molecules in **Scheme 1** have been investigated.

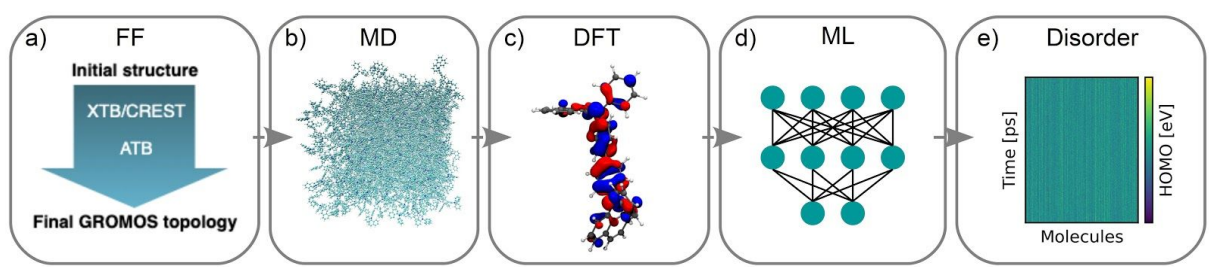

*Figure 1: Computational workflow. The full workflow can be divided into a) conformer search and force field parameterization (FF), b) molecular dynamics simulation (MD), c) density functional theory (DFT) calculations, d) machine learning model (ML) generation, and f) final energetics analysis of the full trajectories to estimate disorder.*

**Molecular dynamics.** For each molecule, the geometry is initially optimized with a conformer search using the submodule CREST of semi-empirical program package XTB[56–58] to find the minimum energy conformation. The resulting structure is submitted to the online platform ATB[59,60], which automatically assigns GROMOS force field parameters and partial charges.[61] For the simulation of the melt structures, we use the molecular dynamics (MD) code LAMMPS.[62] Throughout this work we use a timestep of 1 fs, a cutoff of 14 Å for short-range interactions in combination with the PPPM long-range solver for electrostatics and Nosé-Hoover style thermostating and barostating. The MD simulation is initialized by placing randomly rotated molecules on a grid, followed by a cascade of NPT runs to equilibrate the density. During an initial pressure ramp from 10 to 1 atm and at a temperature of 300 K, the loose initial structure is condensed into a solid over a period of 200 ps. Then the long-range solver is activated followed by an additional run of 200 ps at the final pressure of 1 atm. In order to further equilibrate the density, the system is heated from 300 K to 700 K and subsequently cooled back to 300 K, in each case as linear ramps of 200 ps. At the final conditions of 300 K and 1 atm, the system is equilibrated for another 500 ps. Finally, in the production runs, we extract snapshots in an interval of 1 ps over a period of 5 ns and in an interval of 1 fs over a period of 1 ps. A typical example for the development of the mass density during the equilibration and production runs is shown in Figure S13 of the **SI**.

**DFT.** In order to generate a suitable conformer-energy dataset to train machine learning models, we performed density functional theory (DFT) calculations with the Turbomole 7.4 software package on randomly sampled conformers.[63] We used the resolution of identity approximation with a def2-SV(P) basis set on a M3 grid and the B3-LYP functional.[64] We extracted the total energy as well as vacuum HOMO and LUMO energy levels from DFT single point energy calculations. We simply used the last 20-100 snapshots of the MD trajectories with large timesteps, which leads to a set of less correlated conformers compared to the trajectories with a smaller timestep. Each snapshot has 512 molecules (mCP has 1000 molecules), which leads to approximately 10'000 - 50'000 conformers and corresponding energy levels of each material for training. Additionally, we randomly picked 6 molecules in each simulation box and calculated DFT-energies for the full trajectories from 0 to 5000 ps with a total of approximately 30'000 conformational energies, which serves as an additional test set and to estimate approximate DFT-based disorder values. Moreover, we calculated energy levels using DFT along 6 trajectories for the small timestep MD trajectories of randomly selected molecules. We further experimented with different basis set and grid options (see **SI** for details). There is a small dependence of the DFT results on the basis set size, which likely improves going to def2-TZVP and def2-QZVP, at the cost of a significant increase in computational effort, which makes a DFT reference uncertainty of a few meV seem acceptable.

**ML.** For our machine learning model, we chose a fully connected feed-forward neural network (NN) of 3 layers with a leaky RELU activation function ($\alpha = 0.05$) followed by a single linear classification layer. We trained a separate NN for each material on the respective DFT-energy dataset generated in the previous step. As input features we used rotationally invariant, geometry derived descriptors, such as inverse distances, bond and dihedral angles. Because of the large size of some organic semiconductors in this study, we further restricted the number of inverse distances to a set of close neighbors per atom (including hydrogens). The input features were subsequently concatenated and standardized. The NNs were trained using the Adam optimizer with a learning rate decreasing from $5 \times 10^{-4}$ to $1 \times 10^{-5}$ over 400 epochs. We found a good convergence for a broad set of hyperparameters and selected a NN with 3 hidden layers à 1000 neurons before the final regression layer. To prevent overfitting, we applied L1-regularization for kernel and bias ($10^{-6}$-$10^{-7}$). Note, an exhaustive hyperparameter optimization was not found to be necessary, which is why only the regularization parameters were tuned. For evaluation we choose a validation-training set split ratio of 0.1.

**Dynamic disorder.** As mentioned in the introduction, the conformational energy disorder of molecular solids has dynamic and static contributions. Combined with electrostatic contributions which are described in more detail in prior work,[21,34,65] they add up to the total distribution of energy levels. Provided the energy $U_i(t)$ is given for each molecule $i$ at every point in time by our ML model, the conformational disorder components can be separated as shown in **Figure 3**. This is done by evaluating either the variance of energy levels for a single molecule with time or the distribution of time-averaged energies within the ensemble. The dynamic disorder is characterized by the expectation value of the variance of energy fluctuations of each molecule with time:

$$\sigma_d^2 = E_i[Var_t[U_i(t)]]. \quad (1)$$

Here, $Var_{t/i}$ represents the variance and $E_{t/i}$ denotes the expectation value with respect to time $t$ or ensemble average $i$, respectively. Correspondingly, the static disorder is defined by the variance of the time-averaged expectation values of the energy levels of each molecule:

$$\sigma_s^2 = Var_i[E_t[U_i(t)]]. \quad (2)$$

If both distributions can be approximated independently by normal distributions, the total conformational disorder $\sigma_t^2 = Var_{i,t}[U(t)]$ is given by the quadratic sum of dynamic and static disorder:

$$\sigma_t^2 = \sigma_d^2 + \sigma_s^2. \quad (3)$$

We note that in the following discussion we neglect the electrostatic contribution and focus solely on conformational disorder.

### 3. Results and discussion

We trained our machine learning model on the DFT-generated HOMO-conformer dataset for each relevant molecule in **Figure 2**. HOMO/LUMO energy levels are used here as a proxy for the IP/EA levels which are relevant for charge transport. However, as we do not take polarization effects into account, explicit polarons and thus IP/EA levels are not required in this discussion. The dependence on the number of training samples can be seen from the learning curve in **Figure 2b**, yielding an exponential reduction of the mean absolute validation error on the training set size. For all molecules a validation error of 1 kcal/mol or lower is achieved with 10'000 random conformers from the large timestep trajectories. The ML-predicted HOMO energies are plotted versus the ground truth of the DFT HOMO levels in a scatterplot for NPB in **Figure 2a**. Here, we achieve a mean absolute error (MAE) of 15 meV with an $r^2$-score of 0.97, which is well below chemical precision and the estimated disorder values. We find that NN trained on ensembles of molecules from only a few snapshots as for NPB nicely generalize on the test trajectories of single molecules for longer time scales ($r^2$-score of 0.96). However, the opposite is not true as NNs trained solely on a few trajectories do not correctly predict other trajectories or snapshots (for NPB only $r^2$-score of 0.6). This is most likely due to the fact that for a single trajectory all reachable dynamic conformations are confined around their positional dependent mean and do not visit the full conformational space. The models trained on random snapshots from the large timestep dataset adequately predict the small timestep trajectory, which is depicted in **Figure 2c** for NPB. The small timestep MD simulation is used to investigate oscillations and hopping time scales in **Figure 5**.

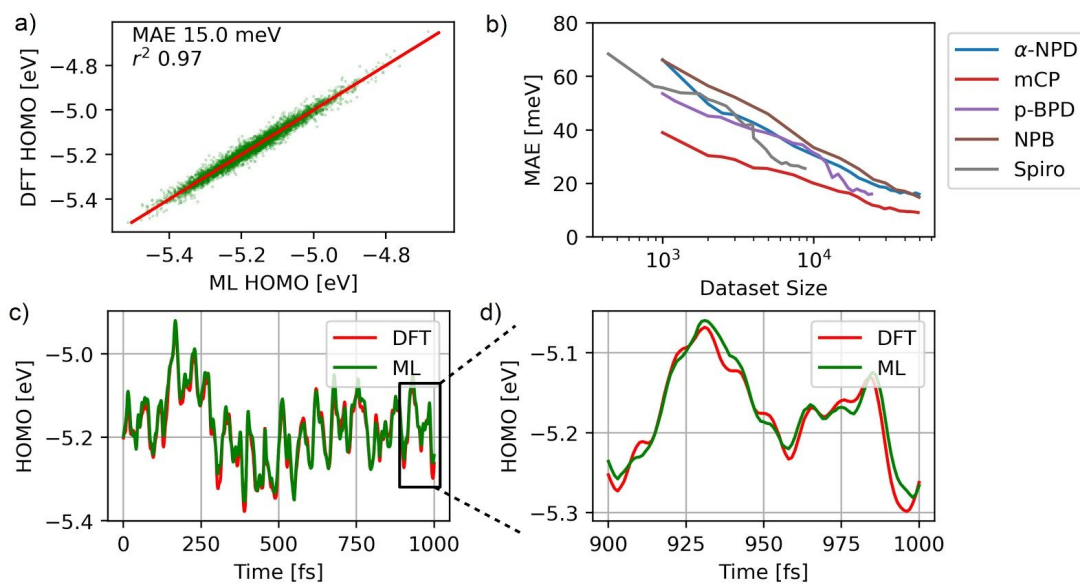

*Figure 2: Machine learning. We trained a NN for all molecules in Scheme 1 to predict HOMO and LUMO levels, respectively. a) Scatter plot of DFT reference values versus predicted validation set for NPB. c) A validation trajectory comparing DFT reference with ML-predicted trajectory for NPB. b) Learning curve for NPD, p-BPD, NPB, Spiro-TAD, and mCP, which shows the mean absolute error (MAE) as a function of training data. d) A zoom into an oscillation of b).*

Applying the trained machine learning model on the full MD trajectories, all conformational energies are predicted, illustrated in **Figure 3a** with NPB as an example. Looking at snapshots in **Figure 3c**, *i.e.,* at all molecules at a specific time, the total conformational disorder can be determined based on the HOMO level distribution and its standard deviation in **Figure 3f**. The total conformational disorder is approximately Gaussian. The HOMO and LUMO level distribution for all molecules is further evaluated directly from DFT reference calculations in **Figure S1** and **S2**. Although the total conformational disorder can be directly inferred from DFT data, a distinction between static and dynamic contributions requires complete trajectories as in **Figure 3b**. In the image of **Figure 3a,** trajectories can be seen as vertical slices along the time axis. The distribution of their mean values is plotted in blue in **Figure 3d**, whereas the scattering around their respective mean values is depicted in **Figure 3e**. Importantly, the machine learning model provides enough trajectories to properly represent the distribution of time-averaged energy levels and their fluctuations. And therefore the standard deviation of the distributions **Figure 3d** and **Figure 3e** can be taken as an estimate for the static and dynamic disorder following Equation 2 and 1, respectively. Additionally, we approximated the static and dynamic disorder based on DFT reference values from six test trajectories in **Table S1** and **S2**.

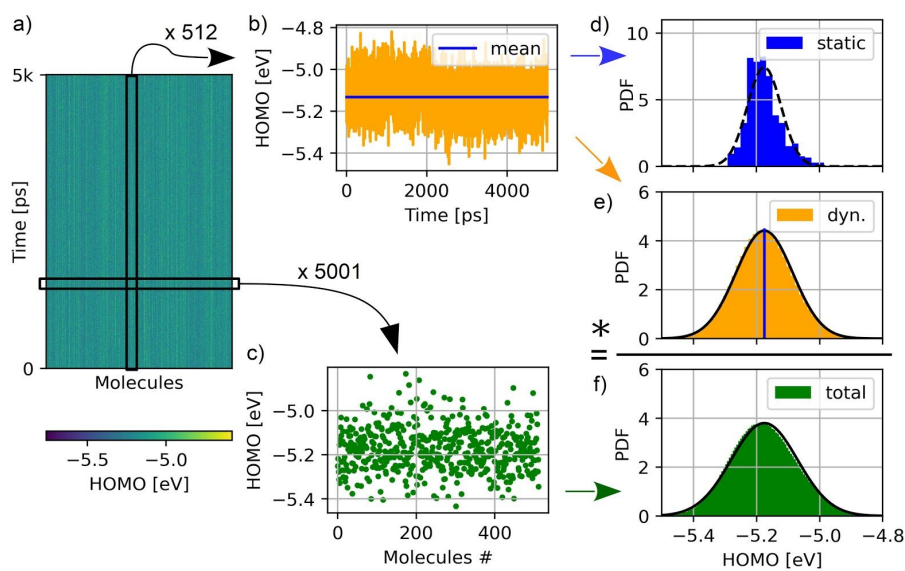

*Figure 3: Definition of disorder. The analysis of static and dynamic energy disorder is illustrated by the example of NPB. a) A heatmap of all NN predictions, where time flows in y-direction. b) One out of 512 trajectories. c) A single snapshot out of 5001 at a specific time showing all molecules. d) A probability density function (PDF) plot of the distribution of all the trajectories mean value representing the static disorder distribution. e) Dynamic contribution of the trajectories. The fluctuations of all trajectories centered to their respective mean value. f) Total conformational disorder distribution from averaged snapshots, which is composed of a convolution of the static and dynamic distributions,* i.e. *the square sum of dynamic and static disorder.*

Following this data analysis we determine disorder components for the HOMO level of all p-type molecules and for the LUMO level of all n-type molecules in **Scheme 1**. The individual disorder values are depicted in **Figure 4** and listed in **Table S3** and **S4** together with the ML validation error and the squared sum of dynamic and static disorder. We find an overall very good agreement of the squared sum with the total conformational disorder value, showing that the disorder is composed of a convolution of independent dynamic and static disorder contributions, which can not be concluded from the six DFT reference trajectories alone due to limited statistics.

In general, we find a notable contribution of the dynamic part to the total conformational disorder. As discussed later, the dynamic disorder has the potential to hinder charge transport.[66] The static disorder seems to be larger for molecules with long rotatable aromatic groups such as TCTA, where the vacuum HOMO level is delocalized and can localize inside groups depending on the conformation (see **Figure S5**). Also, in direct comparison of m-BPD and p-BPD with o-BPD, the more compact and sterically locked ortho-position shows the smallest static disorder. Similarly, with the addition of methyl-groups in the backbone of NPB (yielding a-NPD), static disorder decreases and the ratio between static and dynamic disorder is altered. However, the reduction in static disorder of the methyl-substituted a-NPD is compensated by an increase in dynamic disorder.

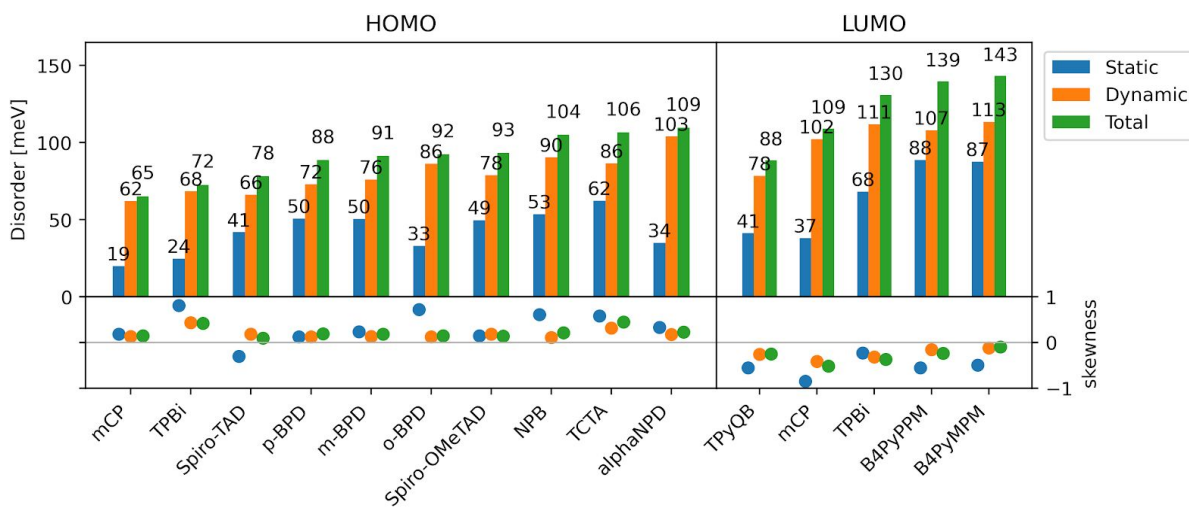

*Figure 4: Disorder components and (a)symmetry.* Energy disorder broken down into dynamic and static parts as illustrated in **Figure 3**. For all n-type molecules the LUMO level and for all p-type molecules the HOMO level is plotted (see **Scheme 1**). All 38.3 million energy levels required for this analysis are generated by our machine learning models. All ML estimates of HOMO and LUMO disorder are given in Table S3, S4 and Figure S8, S9. The corresponding values obtained from DFT reference trajectories with significantly smaller statistics are given in Table S1, S2 for comparison. The lower panels show a measure of skewness of the static, dynamic and total energy level distributions. In particular for the static energy level distributions, we find substantial deviations from symmetrical Gaussian distributions.

While the total and dynamic disorder can be well described with a Gaussian distribution, the static disorder is often slightly asymmetric (see **Figure 3b**, lower panel in **Figure 4**, and SI). Electrons and holes with hopping rates that are lower than the typical fluctuation time of the dynamic disorder will thus experience an asymmetrical distribution of energy levels. Depending on the type of asymmetry, this might lead to a decrease or increase in the concentration of tail states / shallow traps. In the example illustrated in **Figure 3d**, the holes traveling through NPB are in a skewed HOMO distribution that shows more tail states and thus hole traps than what would be expected from a Gaussian distribution of energy levels with the same standard deviation. The opposite behaviour was observed in Friederich *et al.*[55] for hole transport in PEDOT oligomers, where the distribution of HOMO levels showed an extended tail towards lower HOMO energy levels and a sharp edge towards more shallow HOMO energies. One possible explanation of this behavior might be the correlation between total energy and HOMO energy close to the equilibrium conformation. In case of a positive correlation, the HOMO distribution has a global minimum, leading to a distribution as shown in **Figure 3d**, whereas in case of a negative correlation, the opposite asymmetry is observed. This can be seen in the example of PEDOT and in various cases in **Figure 4**, especially for distributions of static LUMO energy levels and for the distribution of static HOMO energy levels of Spiro-TAD.

Depending on the theoretical model used to describe charge transport, also the structurally resolved energy levels of the surroundings of a molecule can be required. For example in kinetic Monte Carlo simulations, charge transport is modeled as stochastic charge transfer between different sites subject to an external electric field. With our machine learning model, the time-resolved energetic neighborhood to a molecule can be explored.

In the following, we will analyse the time resolved energy difference between neighboring molecules at the examples of a small NPB cluster and compare it to typical hopping rates. The power spectrum of the energy level fluctuations of NBP molecules (see **Figure 5d**) shows an approximate 1/f noise-like background, reaching down to a few THz. In addition, we find distinct vibrational modes leading to oscillations with a frequency of ~17 THz (59 fs), which corresponds to the more pronounced oscillations plotted in **Figure 5b**. Furthermore, we find additional frequency components at ~37 and ~51 THz (59 fs and 27 fs, respectively), corresponding to weaker fluctuations that are modulated on the main fluctuations in **Figure 5b**. The frequency of energy level resonances (in terms of Marcus theory[27], *i.e.* taking into account the reorganization energy, see **Eq. 4**) between neighboring molecules is analysed In **Figure 5c**. We find resonances with average occurrence rates of ~56 THz, ~90 THz and ~120 THz (18 fs, 11 fs and 8 fs, respectively), when computing the resonances between a NPB molecule and its nearest 3, 5 and 7 neighbors.

$$W_{ij} = \frac{2\pi |J_{ij}|^2}{h} \sqrt{\frac{\pi}{\lambda k T}} \, e^{-(E_j - E_i + \lambda)^2 / 4\lambda k T} \tag{4}$$

To compare these characteristic fluctuation and resonance times to typical hopping rates, we have to take into account the electronic coupling between NPB molecules, which quantifies the prefactor of the Marcus hoping rate, and which can be interpreted as a hopping attempt frequency. The electronic coupling strongly depends on the intermolecular distance (see **Figure S6**). The pair correlation function of NPB (**Figure 5a**) shows that a substantial amount of NPB molecules have their nearest neighbour at center-of-mass distances between 3 Å and 8 Å (with a stacking motive in which two NPB molecules are situated in close proximity and rotated approximately 90° relative to each other). The next nearest neighbours which are still in direct contact, thus allowing for non-vanishing electronic coupling matrix elements, are located at distances between 9-10 Å.

Electronic couplings $J_{ij}$ and reorganization energies are typically obtained from quantum chemical calculations.[67] For NPB, we obtain $J_{ij} = 0.01 \, eV$ at a distance of ~5 Å and a reorganization energy $\lambda = 0.16 \, eV$, leading to a maximum hopping rate of 4.2 THz for Marcus theory, assuming that the resonance condition is fulfilled, *i.e.* for $E_j - E_i + \lambda = 0$. This translates to a residence time $1/W_{ij}$ of 238 fs, which suggests that the oscillations visible in **Figure 5b** appear averaged for most charge carriers. When taking into account larger hopping distances, the mean electronic coupling is substantially lower, allowing the assumption that the energy levels appear time averaged. This means that the resulting mean energy levels of molecules should be drawn from the static disorder distribution. If we were to remove the distinct oscillations from the frequency spectrum (filter out frequencies higher than 10 THz), the dynamic disorder would shrink by 22 meV from 84 to 62 meV but the average resonance occurrence time increases from 11 fs to 67 fs for 5 neighbours considered.

Nonetheless, we also note that in case of molecules with higher mean electronic couplings than NPB (e.g. $J_{ij} \sim 0.1$ eV, corresponding to a prefactor of the Marcus rate of 418 THz or 2 fs in resonance condition), even distinct dynamic oscillations become comparable or slower than typical hopping rates, leading to a significantly increased effective disorder seen by charge carriers in a classical description. Although this is a classical picture and the analogue is not strictly true, the fast distinct oscillations in **Figure 5b** may be regarded as a way to bring neighbouring energy levels into resonance, or as phonons being absorbed to facilitate hopping as in Marcus theory. We would like to point out that with a ML model introduced into atomistic charge transport models, these relative contributions could be captured and quantified by introducing the time dependent energy levels.

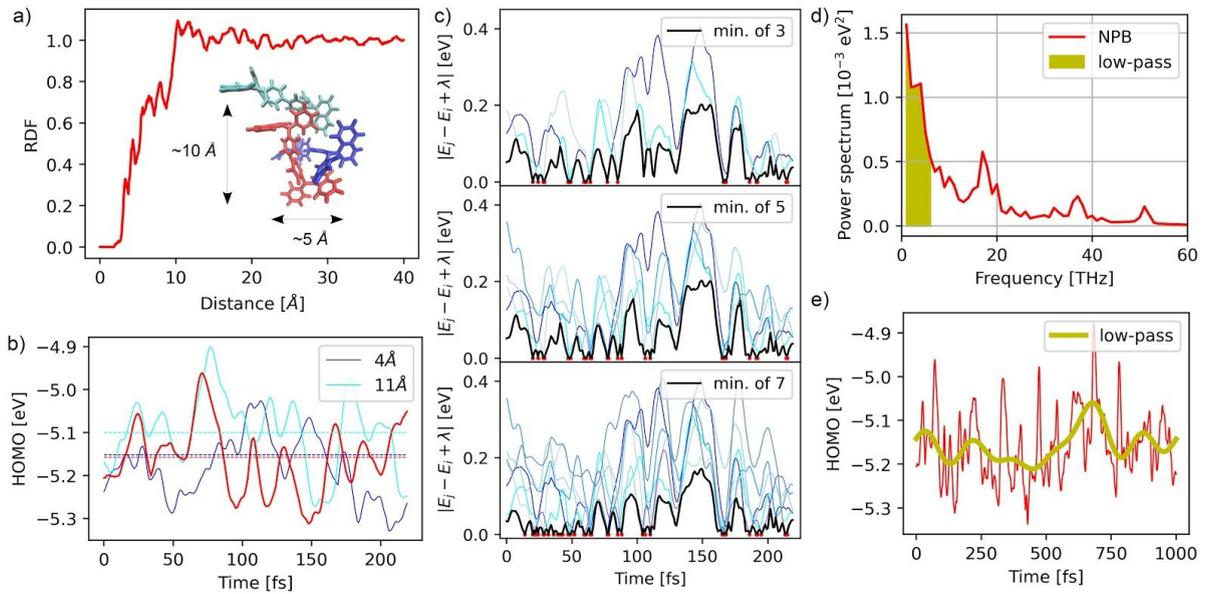

*Figure 5: **Hopping rate and energy matching.** a) Radial distribution or pair correlation function for NPB obtained from averaged MD morphology. The inset shows the stacking of NPB molecules in the amorphous phase with two common length scales as an example arrangement. b) ML prediction of energies of a random NPB molecule (red) and selected neighboring molecules in blue and cyan at ca. 4 and 11 Å distance. c) Absolute energy difference, including reorganization energy $\lambda$, of the central molecule to the 3, 5 and 7 nearest neighbors in shades of blue and their minimum in black, and resonances in red. d) Averaged power spectrum of NPB from FFT of ML predicted trajectories. e) Trajectory of the molecule in b) with low-pass filtered frequencies components in orange.*

For example, a larger grid of realistic time-dependent energy sites could in principle be generated for Monte-Carlo simulations. Also in extensive 3D morphologies, a joint energetic and geometric representation will also reveal possible percolation pathways (see **Figure S4**). This could be explored in future work, integrating ML models in charge transport simulations.

To arrive at a static disorder, which can be used in mesoscopic models, *e.g.,* in kinetic Monte-Carlo simulations, the static conformational disorder computed in this work has to be complemented with the electrostatic disorder $\sigma_p$ which can be contributed to intermolecular electrostatic interaction, *e.g.,* molecular dipole moments. The electrostatic disorder can be calculated with the QuantumPatch method[21,34], which yields $\sigma_p = 70$ meV for NPB. Assuming that electrostatic and static conformational disorder are not correlated, the total static disorder is equal to $\sqrt{\sigma^2_s + \sigma^2_p} = 88\ meV$. This is close to the disorder parameter (87 meV) used in mesoscopic models of charge transport [68–70] that is used by a top-down approach to reproduce the experimentally observed hole mobility of NPB. Moreover, the hole mobility of NPB computed with the kMC model with a realistic morphology and assuming a disorder parameter 0.93 eV yields a hole mobility of 2x10$^{-3}$ cm$^2$/Vs [71] close to experimental value (between 2x10$^{-3}$ cm$^2$/Vs and 3x10$^{-3}$ cm$^2$/Vs according to Chen et al.[72] and Tse et al.[73]). This shows the high relevance of the developed method to accurately compute charge carriers mobility in disordered organic semiconductors.

## 4. Conclusions

In summary we have investigated static and dynamic conformational disorder for various device-relevant organic semiconductors by using machine learning models integrated in multiscale modeling simulations. The ML models were trained using DFT energies to predict the time- and space resolved energy level distribution of complete MD trajectories of disordered thin films. Thereby, static and dynamic disorder contributions were analysed using ~38 million energy level predictions by the ML model, a number that would not have been feasible using quantum mechanical calculations.

We firstly find that static disorder distributions of multiple relevant materials are highly asymmetrical. Consequently the density of states either shows deep tail states in the HOMO-LUMO gap which is unfavourable for charge transport (*e.g.* the HOMO distribution of NPB and TPBi), or it shows a steep decline towards the HOMO-LUMO gap which avoids trapping and thus favours charge transport (*e.g.* the HOMO distribution of Spiro-TAD). We suspect that the underlying mechanism is related to the HOMO energy surface close to local minima of the potential energy surface. A closer study of this mechanism might lead to new molecular design principles.

We secondly find that characteristic oscillation times of molecular energy levels are below maximally possible Marcus hopping rates in disordered materials like NPB, suggesting that dynamic disorder averages out at typical electronic couplings of disordered materials, leaving static disorder as the key contribution to charge transport. Thirdly, we observed that the amount of static disorder is likely correlated to the size and the amount of rotatable side groups (*e.g.* TCTA and BPD), provided that the respective molecular energy level is sufficiently delocalized over the molecule and therefore affected by conformational changes. This effect offers the opportunity to be used as a design rule for future organic semiconductors.

In summary, we showed that the integration of ML models as fast and accurate predictors in multiscale modeling workflows offers the opportunity to study a wide range of interesting and application relevant phenomena, *e.g.* time resolved energy level fluctuations, dynamic disorder and asymmetries in static energy level distributions, which were not accessible before due to prohibitively high computational costs.

**Acknowledgement**

W.W. and M.K. acknowledge the financial support by the Deutsche Forschungsgemeinschaft (DFG) under Germany Excellence Strategy via the Excellence Cluster 3D Matter Made to Order (MMTO EXC-2082, 1–390761711) and the Research Training Group (GRK 2450). P.F. acknowledges funding from the European Union's Horizon 2020 research and innovation programme under the Marie Sklodowska-Curie grant agreement No 795206. We thank Simon Kaiser for kindly providing electronic couplings for NPB. This work was performed on the supercomputer ForHLR II funded by the Ministry of Science, Research and the Arts Baden-Württemberg and by the Federal Ministry of Education and Research.

**References**

(1) Geffroy, B.; Roy, P. le; Prat, C. Organic Light-Emitting Diode (OLED) Technology: Materials, Devices and Display Technologies. *Polym. Int.* **2006**, *55* (6), 572–582. https://doi.org/10.1002/pi.1974.
(2) Zhang, G.; Zhao, J.; Chow, P. C. Y.; Jiang, K.; Zhang, J.; Zhu, Z.; Zhang, J.; Huang, F.; Yan, H. Nonfullerene Acceptor Molecules for Bulk Heterojunction Organic Solar Cells. *Chem. Rev.* **2018**, *118* (7), 3447–3507. https://doi.org/10.1021/acs.chemrev.7b00535.
(3) Kuehne, A. J. C.; Gather, M. C. Organic Lasers: Recent Developments on Materials, Device Geometries, and Fabrication Techniques. *Chem. Rev.* **2016**, *116* (21), 12823–12864. https://doi.org/10.1021/acs.chemrev.6b00172.
(4) Katz, H. E.; Bao, Z.; Gilat, S. L. Synthetic Chemistry for Ultrapure, Processable, and High-Mobility Organic Transistor Semiconductors. *Acc. Chem. Res.* **2001**, *34* (5), 359–369. https://doi.org/10.1021/ar990114j.
(5) Diao, Y.; Tee, B. C.-K.; Giri, G.; Xu, J.; Kim, D. H.; Becerril, H. A.; Stoltenberg, R. M.; Lee, T. H.; Xue, G.; Mannsfeld, S. C. B.; Bao, Z. Solution Coating of Large-Area Organic Semiconductor Thin Films with Aligned Single-Crystalline

(6) Irimia-Vladu, M. "Green" Electronics: Biodegradable and Biocompatible Materials and Devices for Sustainable Future. *Chem. Soc. Rev.* **2014**, *43* (2), 588–610. https://doi.org/10.1039/C3CS60235D.
(7) Fediai, A.; Symalla, F.; Friederich, P.; Wenzel, W. Disorder Compensation Controls Doping Efficiency in Organic Semiconductors. *Nat. Commun.* **2019**, *10* (1), 4547. https://doi.org/10.1038/s41467-019-12526-6.
(8) Li, J.; Duchemin, I.; Maria Roscioni, O.; Friederich, P.; Anderson, M.; Como, E. D.; Kociok-Köhn, G.; Wenzel, W.; Zannoni, C.; Beljonne, D.; Blase, X.; D'Avino, G. Host Dependence of the Electron Affinity of Molecular Dopants. *Mater. Horiz.* **2019**, *6* (1), 107–114. https://doi.org/10.1039/C8MH00921J.
(9) Reiser, P.; Müller, L.; Sivanesan, V.; Lovrincic, R.; Barlow, S.; Marder, S. R.; Pucci, A.; Jaegermann, W.; Mankel, E.; Beck, S. Dopant Diffusion in Sequentially Doped Poly(3-Hexylthiophene) Studied by Infrared and Photoelectron Spectroscopy. *J. Phys. Chem. C* **2018**, *122* (26), 14518–14527. https://doi.org/10.1021/acs.jpcc.8b02657.
(10) Tyagi, P.; Tuli, S.; Srivastava, R. Study of Fluorescence Quenching Due to 2, 3, 5, 6-Tetrafluoro-7, 7′, 8, 8′-Tetracyano Quinodimethane and Its Solid State Diffusion Analysis Using Photoluminescence Spectroscopy. *J. Chem. Phys.* **2015**, *142* (5), 054707. https://doi.org/10.1063/1.4907274.
(11) Yuan, Y.; Giri, G.; Ayzner, A. L.; Zoombelt, A. P.; Mannsfeld, S. C. B.; Chen, J.; Nordlund, D.; Toney, M. F.; Huang, J.; Bao, Z. Ultra-High Mobility Transparent Organic Thin Film Transistors Grown by an off-Centre Spin-Coating Method. *Nat. Commun.* **2014**, *5* (1), 3005. https://doi.org/10.1038/ncomms4005.
(12) Chu, M.; Fan, J.-X.; Yang, S.; Liu, D.; Ng, C. F.; Dong, H.; Ren, A.-M.; Miao, Q. Halogenated Tetraazapentacenes with Electron Mobility as High as 27.8 Cm2 V−1 S−1 in Solution-Processed n-Channel Organic Thin-Film Transistors. *Adv. Mater.* **2018**, *30* (38), 1803467. https://doi.org/10.1002/adma.201803467.
(13) Helfrich, W.; Schneider, W. G. Recombination Radiation in Anthracene Crystals. *Phys. Rev. Lett.* **1965**, *14* (7), 229–231. https://doi.org/10.1103/PhysRevLett.14.229.
(14) Bässler, H. Charge Transport in Disordered Organic Photoconductors a Monte Carlo Simulation Study. *Phys. Status Solidi B* **1993**, *175* (1), 15–56. https://doi.org/10.1002/pssb.2221750102.
(15) Bässler, H.; Köhler, A. 7 - Charge Carrier Mobility in Amorphous Organic Semiconductors. In *Organic Light-Emitting Diodes (OLEDs)*; Buckley, A., Ed.; Woodhead Publishing Series in Electronic and Optical Materials; Woodhead Publishing, 2013; pp 192–234. https://doi.org/10.1533/9780857098948.2.192.
(16) Bässler, H.; Köhler, A. Charge Transport in Organic Semiconductors. In *Unimolecular and Supramolecular Electronics I: Chemistry and Physics Meet at Metal-Molecule Interfaces*; Metzger, R. M., Ed.; Topics in Current Chemistry; Springer: Berlin, Heidelberg, 2012; pp 1–65. https://doi.org/10.1007/128_2011_218.
(17) Friederich, P.; Fediai, A.; Kaiser, S.; Konrad, M.; Jung, N.; Wenzel, W. Toward Design of Novel Materials for Organic Electronics. *Adv. Mater.* **2019**, *31* (26), 1808256. https://doi.org/10.1002/adma.201808256.
(18) Friederich, P.; Konrad, M.; Strunk, T.; Wenzel, W. Machine Learning of Correlated Dihedral Potentials for Atomistic Molecular Force Fields. *Sci. Rep.* **2018**, *8* (1), 2559. https://doi.org/10.1038/s41598-018-21070-0.
(19) Friederich, P.; Gómez, V.; Sprau, C.; Meded, V.; Strunk, T.; Jenne, M.; Magri, A.; Symalla, F.; Colsmann, A.; Ruben, M.; Wenzel, W. Rational In Silico Design of an Organic Semiconductor with Improved Electron Mobility. *Adv. Mater.* **2017**, *29* (43), 1703505. https://doi.org/10.1002/adma.201703505.
(20) Symalla, F.; Friederich, P.; Massé, A.; Meded, V.; Coehoorn, R.; Bobbert, P.; Wenzel, W. Charge Transport by Superexchange in Molecular Host-Guest Systems. *Phys. Rev. Lett.* **2016**, *117* (27), 276803. https://doi.org/10.1103/PhysRevLett.117.276803.
(21) Friederich, P.; Meded, V.; Symalla, F.; Elstner, M.; Wenzel, W. QM/QM Approach to Model Energy Disorder in Amorphous Organic Semiconductors. *J. Chem. Theory Comput.* **2015**, *11* (2), 560–567. https://doi.org/10.1021/ct501023n.
(22) Andrienko, D. Multiscale Concepts in Simulations of Organic Semiconductors. In *Handbook of Materials Modeling: Methods: Theory and Modeling*; Andreoni, W., Yip, S., Eds.; Springer International Publishing: Cham, 2020; pp 1431–1442. https://doi.org/10.1007/978-3-319-44677-6_39.
(23) Fediai, A.; Emering, A.; Symalla, F.; Wenzel, W. Disorder-Driven Doping Activation in Organic Semiconductors. *Phys. Chem. Chem. Phys.* **2020**, *22* (18), 10256–10264. https://doi.org/10.1039/D0CP01333A.
(24) Symalla, F.; Fediai, A.; Armleder, J.; Kaiser, S.; Strunk, T.; Neumann, T.; Wenzel, W. 43-3: Ab-Initio Simulation of Doped Injection Layers. *SID Symp. Dig. Tech. Pap.* **2020**, *51* (1), 630–633. https://doi.org/10.1002/sdtp.13946.
(25) Groves, C. Simulating Charge Transport in Organic Semiconductors and Devices: A Review. *Rep. Prog. Phys.* **2016**, *80* (2), 026502. https://doi.org/10.1088/1361-6633/80/2/026502.
(26) Miller, A.; Abrahams, E. Impurity Conduction at Low Concentrations. *Phys. Rev.* **1960**, *120* (3), 745–755. https://doi.org/10.1103/PhysRev.120.745.
(27) Marcus, R. A. Electron Transfer Reactions in Chemistry. Theory and Experiment. *Rev. Mod. Phys.* **1993**, *65* (3), 599–610. https://doi.org/10.1103/RevModPhys.65.599.
(28) Pasveer, W. F.; Cottaar, J.; Tanase, C.; Coehoorn, R.; Bobbert, P. A.; Blom, P. W. M.; de Leeuw, D. M.; Michels, M. A. J. Unified Description of Charge-Carrier Mobilities in Disordered Semiconducting Polymers. *Phys. Rev. Lett.* **2005**, *94* (20), 206601. https://doi.org/10.1103/PhysRevLett.94.206601.
(29) Bouhassoune, M.; Mensfoort, S. L. M. van; Bobbert, P. A.; Coehoorn, R. Carrier-Density and Field-Dependent Charge-Carrier Mobility in Organic Semiconductors with Correlated Gaussian Disorder. *Org. Electron.* **2009**, *10* (3), 437–445. https://doi.org/10.1016/j.orgel.2009.01.005.
(30) Massé, A.; Friederich, P.; Symalla, F.; Liu, F.; Nitsche, R.; Coehoorn, R.; Wenzel, W.; Bobbert, P. A. Ab Initio Charge-Carrier Mobility Model for Amorphous Molecular Semiconductors. *Phys. Rev. B* **2016**, *93* (19), 195209. https://doi.org/10.1103/PhysRevB.93.195209.
(31) Kirkpatrick, J.; Marcon, V.; Nelson, J.; Kremer, K.; Andrienko, D. Charge Mobility of Discotic Mesophases: A Multiscale Quantum and Classical Study. *Phys. Rev. Lett.* **2007**, *98* (22), 227402. https://doi.org/10.1103/PhysRevLett.98.227402.
(32) Rodin, V.; Symalla, F.; Meded, V.; Friederich, P.; Danilov, D.; Poschlad, A.; Nelles, G.; von Wrochem, F.; Wenzel, W. Generalized Effective-Medium Model for the Carrier Mobility in Amorphous Organic Semiconductors. *Phys. Rev. B* **2015**, *91* (15), 155203. https://doi.org/10.1103/PhysRevB.91.155203.
(33) Rühle, V.; Lukyanov, A.; May, F.; Schrader, M.; Vehoff, T.; Kirkpatrick, J.; Baumeier, B.; Andrienko, D. Microscopic Simulations of Charge Transport in Disordered Organic Semiconductors. *J. Chem. Theory Comput.* **2011**, *7* (10), 3335–3345. https://doi.org/10.1021/ct200388s.
(34) Friederich, P.; Meded, V.; Poschlad, A.; Neumann, T.; Rodin, V.; Stehr, V.; Symalla, F.; Danilov, D.; Lüdemann, G.; Fink, R. F.; Kondov, I.; Wrochem, F. von; Wenzel, W. Molecular Origin of the Charge Carrier Mobility in Small Molecule

(35) Poelking, C.; Tietze, M.; Elschner, C.; Olthof, S.; Hertel, D.; Baumeier, B.; Würthner, F.; Meerholz, K.; Leo, K.; Andrienko, D. Impact of Mesoscale Order on Open-Circuit Voltage in Organic Solar Cells. *Nat. Mater.* **2015**, *14* (4), 434–439. https://doi.org/10.1038/nmat4167.
(36) Yavuz, I.; Martin, B. N.; Park, J.; Houk, K. N. Theoretical Study of the Molecular Ordering, Paracrystallinity, And Charge Mobilities of Oligomers in Different Crystalline Phases. *J. Am. Chem. Soc.* **2015**, *137* (8), 2856–2866. https://doi.org/10.1021/ja5076376.
(37) Hervet, H.; Dianoux, A. J.; Lechner, R. E.; Volino, F. Neutron Scattering Study of Methyl Group Rotation in Solid Para-Azoxyanisole (PAA). *J. Phys.* **1976**, *37* (5), 587–594. https://doi.org/10.1051/jphys:01976003705058700.
(38) Harada, J.; Ogawa, K. Invisible but Common Motion in Organic Crystals: A Pedal Motion in Stilbenes and Azobenzenes. *J. Am. Chem. Soc.* **2001**, *123* (44), 10884–10888. https://doi.org/10.1021/ja011197d.
(39) Llamas-Saiz, A. L.; Foces-Foces, C.; Fontenas, C.; Jagerovic, N.; Elguero, J. The Search for Proton Mobility in Solid Pyrazoles: Molecular and Crystal Structure of 3(5)-Phenyl-4-Bromo-5(3)-Methylpyrazole. *J. Mol. Struct.* **1999**, *484* (1), 197–205. https://doi.org/10.1016/S0022-2860(98)00905-3.
(40) White, M. A.; Wasylishen, R. E.; Eaton, P. E.; Xiong, Y.; Pramod, K.; Nodari, N. Orientational Disorder in Solid Cubane: A Thermodynamic and Carbon-13 NMR Study. *J. Phys. Chem.* **1992**, *96* (1), 421–425. https://doi.org/10.1021/j100180a078.
(41) Cruz-Cabeza, A. J.; Day, G. M.; Jones, W. Structure Prediction, Disorder and Dynamics in a DMSO Solvate of Carbamazepine. *Phys. Chem. Chem. Phys.* **2011**, *13* (28), 12808–12816. https://doi.org/10.1039/C1CP20927B.
(42) Richards, T.; Bird, M.; Sirringhaus, H. A Quantitative Analytical Model for Static Dipolar Disorder Broadening of the Density of States at Organic Heterointerfaces. *J. Chem. Phys.* **2008**, *128* (23), 234905. https://doi.org/10.1063/1.2937729.
(43) Tummala, N. R.; Zheng, Z.; Aziz, S. G.; Coropceanu, V.; Brédas, J.-L. Static and Dynamic Energetic Disorders in the C60, PC61BM, C70, and PC71BM Fullerenes. *J. Phys. Chem. Lett.* **2015**, *6* (18), 3657–3662. https://doi.org/10.1021/acs.jpclett.5b01709.
(44) Zheng, Z.; Tummala, N. R.; Wang, T.; Coropceanu, V.; Brédas, J.-L. Charge-Transfer States at Organic–Organic Interfaces: Impact of Static and Dynamic Disorders. *Adv. Energy Mater.* **2019**, *9* (14), 1803926. https://doi.org/10.1002/aenm.201803926.
(45) Kupgan, G.; Chen, X.-K.; Brédas, J.-L. Low Energetic Disorder in Small-Molecule Non-Fullerene Electron Acceptors. *ACS Mater. Lett.* **2019**, *1* (3), 350–353. https://doi.org/10.1021/acsmaterialslett.9b00248.
(46) Vehoff, T.; Chung, Y. S.; Johnston, K.; Troisi, A.; Yoon, D. Y.; Andrienko, D. Charge Transport in Self-Assembled Semiconducting Organic Layers: Role of Dynamic and Static Disorder. *J. Phys. Chem. C* **2010**, *114* (23), 10592–10597. https://doi.org/10.1021/jp101738g.
(47) Vehoff, T.; Baumeier, B.; Troisi, A.; Andrienko, D. Charge Transport in Organic Crystals: Role of Disorder and Topological Connectivity. *J. Am. Chem. Soc.* **2010**, *132* (33), 11702–11708. https://doi.org/10.1021/ja104380c.
(48) McMahon, D. P.; Troisi, A. Organic Semiconductors: Impact of Disorder at Different Timescales. *ChemPhysChem* **2010**, *11* (10), 2067–2074. https://doi.org/10.1002/cphc.201000182.
(49) Ishii, H.; Kobayashi, N.; Hirose, K. Carrier Transport Calculations of Organic Semiconductors with Static and Dynamic Disorder. *Jpn. J. Appl. Phys.* **2019**, *58* (11), 110501. https://doi.org/10.7567/1347-4065/ab4b61.
(50) Rühle, V.; Junghans, C.; Lukyanov, A.; Kremer, K.; Andrienko, D. Versatile Object-Oriented Toolkit for Coarse-Graining Applications. *J. Chem. Theory Comput.* **2009**, *5* (12), 3211–3223. https://doi.org/10.1021/ct900369w.
(51) Schleder, G. R.; Padilha, A. C. M.; Acosta, C. M.; Costa, M.; Fazzio, A. From DFT to Machine Learning: Recent Approaches to Materials Science–a Review. *J. Phys. Mater.* **2019**, *2* (3), 032001. https://doi.org/10.1088/2515-7639/ab084b.
(52) Boese, R.; Antipin, M. Yu.; Bläser, D.; Lyssenko, K. A. Molecular Crystal Structure of Acetylacetone at 210 and 110 K: Is the Crystal Disorder Static or Dynamic? *J. Phys. Chem. B* **1998**, *102* (44), 8654–8660. https://doi.org/10.1021/jp980121+.
(53) Troisi, A.; Cheung, D. L. Transition from Dynamic to Static Disorder in One-Dimensional Organic Semiconductors. *J. Chem. Phys.* **2009**, *131* (1), 014703. https://doi.org/10.1063/1.3167406.
(54) Wang, L.; Li, Q.; Shuai, Z.; Chen, L.; Shi, Q. Multiscale Study of Charge Mobility of Organic Semiconductor with Dynamic Disorders. *Phys. Chem. Chem. Phys.* **2010**, *12* (13), 3309–3314. https://doi.org/10.1039/B913183C.
(55) Friederich, P.; Leon, S.; Ospina, J. D. P.; Roch, L.; Aspuru-Guzik, A. The Influence of Sorbitol Doping on Aggregation and Electronic Properties of PEDOT:PSS: A Theoretical Study. *Mach. Learn. Sci. Technol.* **2020**. https://doi.org/10.1088/2632-2153/ab983b.
(56) Bannwarth, C.; Caldeweyher, E.; Ehlert, S.; Hansen, A.; Pracht, P.; Seibert, J.; Spicher, S.; Grimme, S. Extended Tight-Binding Quantum Chemistry Methods. *WIREs Comput. Mol. Sci. n/a* (n/a), e01493. https://doi.org/10.1002/wcms.1493.
(57) Pracht, P.; Bohle, F.; Grimme, S. Automated Exploration of the Low-Energy Chemical Space with Fast Quantum Chemical Methods. *Phys. Chem. Chem. Phys.* **2020**, *22* (14), 7169–7192. https://doi.org/10.1039/C9CP06869D.
(58) Grimme, S. Exploration of Chemical Compound, Conformer, and Reaction Space with Meta-Dynamics Simulations Based on Tight-Binding Quantum Chemical Calculations. *J. Chem. Theory Comput.* **2019**, *15* (5), 2847–2862. https://doi.org/10.1021/acs.jctc.9b00143.
(59) Malde, A. K.; Zuo, L.; Breeze, M.; Stroet, M.; Poger, D.; Nair, P. C.; Oostenbrink, C.; Mark, A. E. An Automated Force Field Topology Builder (ATB) and Repository: Version 1.0. *J. Chem. Theory Comput.* **2011**, *7* (12), 4026–4037. https://doi.org/10.1021/ct200196m.
(60) Stroet, M.; Caron, B.; Visscher, K. M.; Geerke, D. P.; Malde, A. K.; Mark, A. E. Automated Topology Builder Version 3.0: Prediction of Solvation Free Enthalpies in Water and Hexane. *J. Chem. Theory Comput.* **2018**, *14* (11), 5834–5845. https://doi.org/10.1021/acs.jctc.8b00768.
(61) Schmid, N.; Eichenberger, A. P.; Choutko, A.; Riniker, S.; Winger, M.; Mark, A. E.; van Gunsteren, W. F. Definition and Testing of the GROMOS Force-Field Versions 54A7 and 54B7. *Eur. Biophys. J.* **2011**, *40* (7), 843. https://doi.org/10.1007/s00249-011-0700-9.
(62) Plimpton, S. Fast Parallel Algorithms for Short-Range Molecular Dynamics. *J. Comput. Phys.* **1995**, *117* (1), 1–19. https://doi.org/10.1006/jcph.1995.1039.
(63) Balasubramani, S. G.; Chen, G. P.; Coriani, S.; Diedenhofen, M.; Frank, M. S.; Franzke, Y. J.; Furche, F.; Grotjahn, R.; Harding, M. E.; Hättig, C.; Hellweg, A.; Helmich-Paris, B.; Holzer, C.; Huniar, U.; Kaupp, M.; Marefat Khah, A.; Karbalaei

Khani, S.; Müller, T.; Mack, F.; Nguyen, B. D.; Parker, S. M.; Perlt, E.; Rappoport, D.; Reiter, K.; Roy, S.; Rückert, M.; Schmitz, G.; Sierka, M.; Tapavicza, E.; Tew, D. P.; van Wüllen, C.; Voora, V. K.; Weigend, F.; Wodyński, A.; Yu, J. M. TURBOMOLE: Modular Program Suite for Ab Initio Quantum-Chemical and Condensed-Matter Simulations. *J. Chem. Phys.* **2020**, *152* (18), 184107. https://doi.org/10.1063/5.0004635.

(64) Kim, K.; Jordan, K. D. Comparison of Density Functional and MP2 Calculations on the Water Monomer and Dimer. *J. Phys. Chem.* **1994**, *98* (40), 10089–10094. https://doi.org/10.1021/j100091a024.

(65) Friederich, P.; Symalla, F.; Meded, V.; Neumann, T.; Wenzel, W. Ab Initio Treatment of Disorder Effects in Amorphous Organic Materials: Toward Parameter Free Materials Simulation. *J. Chem. Theory Comput.* **2014**, *10* (9), 3720–3725. https://doi.org/10.1021/ct500418f.

(66) Illig, S.; Eggeman, A. S.; Troisi, A.; Jiang, L.; Warwick, C.; Nikolka, M.; Schweicher, G.; Yeates, S. G.; Henri Geerts, Y.; Anthony, J. E.; Sirringhaus, H. Reducing Dynamic Disorder in Small-Molecule Organic Semiconductors by Suppressing Large-Amplitude Thermal Motions. *Nat. Commun.* **2016**, *7* (1), 10736. https://doi.org/10.1038/ncomms10736.

(67) Glaeser, R. M.; Berry, R. S. Mobilities of Electrons and Holes in Organic Molecular Solids. Comparison of Band and Hopping Models. *J. Chem. Phys.* **1966**, *44* (10), 3797–3810. https://doi.org/10.1063/1.1726537.

(68) Massé, A.; Friederich, P.; Symalla, F.; Liu, F.; Meded, V.; Coehoorn, R.; Wenzel, W.; Bobbert, P. A. Effects of Energy Correlations and Superexchange on Charge Transport and Exciton Formation in Amorphous Molecular Semiconductors: An Ab Initio Study. *Phys. Rev. B* **2017**, *95* (11), 115204. https://doi.org/10.1103/PhysRevB.95.115204.

(69) Liu, F.; Massé, A.; Friederich, P.; Symalla, F.; Nitsche, R.; Wenzel, W.; Coehoorn, R.; Bobbert, P. A. Ab Initio Modeling of Steady-State and Time-Dependent Charge Transport in Hole-Only α-NPD Devices. *Appl. Phys. Lett.* **2016**, *109* (24), 243301. https://doi.org/10.1063/1.4971969.

(70) de Vries, X.; Friederich, P.; Wenzel, W.; Coehoorn, R.; Bobbert, P. A. Full Quantum Treatment of Charge Dynamics in Amorphous Molecular Semiconductors. *Phys. Rev. B* **2018**, *97* (7), 075203. https://doi.org/10.1103/PhysRevB.97.075203.

(71) Friederich, P.; Fediai, A.; Li, J.; Mondal, A.; Kotadiya, N. B.; Symalla, F.; Wetzelaer, G.-J. A. H.; Andrienko, D.; Blase, X.; Beljonne, D.; Blom, P. W. M.; Brédas, J.-L.; Wenzel, W. The Influence of Impurities on the Charge Carrier Mobility of Small Molecule Organic Semiconductors. *ArXiv190811854 Cond-Mat Phys.* **2020**.

(72) Chen, B.; Lee, C.; Lee, S.; Webb, P.; Chan, Y.; Gambling, W.; Tian, H.; Zhu, W. Improved Time-of-Flight Technique for Measuring Carrier Mobility in Thin Films of Organic Electroluminescent Materials. *Jpn. J. Appl. Phys.* **2000**, *39* (3R), 1190. https://doi.org/10.1143/JJAP.39.1190.

(73) Tse, S. C.; Tsang, S. W.; So, S. K. Polymeric Conducting Anode for Small Organic Transporting Molecules in Dark Injection Experiments. *J. Appl. Phys.* **2006**, *100* (6), 063708. https://doi.org/10.1063/1.2348640.